\documentstyle[11pt,a4]{article}
\def\be{\nopagebreak[3]\begin{equation}}
\def\ee{\end{equation}}
\def\ba{\nopagebreak[3]\begin{eqnarray}}
\def\ea{\end{eqnarray}}
\def\nl{\nonumber \\}
\def\ni{\noindent}
\newcommand{\scs}[1]{{\scriptscriptstyle #1}}
\begin{document}
\begin{titlepage}
\begin{flushright}
 NBI-HE-92-62\\
 September 1992
\end{flushright}
\vspace*{36pt}
\begin{center}
{\Large \bf
LOCAL SYMMETRY IN THE KAZAKOV-MIGDAL GAUGE MODEL}
\end{center}
\vspace{36pt}
\begin{center}
 {\large D.V. Boulatov}\\
\vspace{12pt}
{\em The Niels Bohr Institute\\
University of Copenhagen\\
Blegdamsvej 17, 2100 Copenhagen \O \\
Denmark}
\vspace{36pt}
\end{center}
\begin{center}
{\bf Abstract}
\end{center}

The spectrum of observables in the induced lattice gauge model proposed
recently by V.A.Kazakov and A.A.Migdal obeys the local-confinement
selection rule. The underlying local continuous symmetry cannot be
spontaneously broken within the model.
\vfill
\end{titlepage}

\newpage

\vspace*{36pt}
V.Kazakov and A.Migdal have recently proposed an induced lattice gauge
model which could (in a continuum limit) be equivalent to QCD
\cite{KM}. The gauge self-interaction is induced by
the hermitean-matrix-valued scalar field,
$\Phi_{ab}(x)=\overline{\Phi}_{ba}(x),\ a,b=1,\ldots,N$, coupled to link
gauge variables, $U_{\mu}(x)\in SU(N)$. $x$ is an index numbering sites
of a regular $D$-dimensional lattice; $\mu=1,\ldots,D$ are $D$
directions and $U_{-\mu}(x)=U^+_{\mu}(x-\mu)$.
The lattice action can be written in the form

\be
S=-N\sum_{x,\mu}tr\Phi(x)U_{\mu}(x)\Phi(x+\mu)U_{\mu}^+(x)\ +\
\sum_{x}trV(\Phi(x))
\label{action}
\ee

\ni
where $V(z)$ is a general potential (one could add a kinetic term for
$\Phi(x)$ as well). This interesting model is soluble at large $N$ by the
mean field technique \cite{Mig,Gross,KhMak,Mak}. Gauge variables are
independent and can be integrated out at each link by the Itzykson-Zuber
formula \cite{ItZub}

\be
\int  dU \: e^{Ntr\phi U\psi U^+} = N^{-N(N-1)/2}
\prod_{n=1}^{N-1}n!\frac{\det_{ab}
e^{ N\phi_a\psi_b}}{\Delta(\phi)\Delta(\psi)}
\label{ItZub}
\ee

\ni
where $\phi$ and $\psi$ can be regarded without loss of generality as
real diagonal matrices; $\Delta(\phi)$ is the Van-der-Monde determinant.
As a result, one gets theory where the role of dynamical variables is
played by $N$ eigenvalues of a matrix $\Phi(x)$. In the $N\to\infty$
limit the mean field approximation gives an exact answer.

As the scalar fields are auxiliary, the complete set of observables in
the model is given by the correlators of Wilson loops

\be
A(L_1,\ldots,L_k)=\langle
\chi_{\scs{R_1}}(\prod_{L_1}U)\chi_{\scs{R_2}}(\prod_{L_2}U)
\;\ldots\;\chi_{\scs{R_k}}(\prod_{L_k}U)\rangle
\label{corr}
\ee

\ni
where $\chi_{\scs{R_i}}(\prod_{L_i}U)$ is the character of an $R_i$'th
irrep of $SU(N)$ as the function of the product of gauge variables along
an $i$'th loop, $L_i$. In general, loops can be (self)intersecting. As
characters are traces of matrix elements, we can, using
Clebsch-Gordan coefficients, re-expand eq. (\ref{corr}) at links
of intersections getting the following basic integral to calculate

\be
I_{ab}^R(\phi,\psi)=\int  dU \:D^R_{ab}(U)\: e^{Ntr\phi U\psi U^+}
\label{int}
\ee

\ni
where $D^R_{ab}(U)$ is a matrix element of an irrep $R$;
$a,b=1,\ldots,d_R$ are indices in a space, $V_R$, of R
($d_R$ is the dimension of $V_R$). $\phi$ and $\psi$ are real and diagonal
without loss of generality, since the field $\Phi(x)$ can be
diagonalized independently at each link by a pure gauge transformation.

The one-link action, $tr\phi U\psi U^+$, in eq. (\ref{int})
is symmetric with respect to the multiplication by
$U(1)^{\otimes N}$ matrices on the left and on the right:

\be
U_{ab}\to U_{ab}e^{i(\alpha_a+\beta_b)}
\label{trans}
\ee

\ni
Consequenses of this symmetry were thoroughly investigated in the contex
of $D=1$ compactified string theory in ref. \cite{BK}. It gives rise to
the following selection rule for representations

\be
I^R_{ab}(\phi,\psi)=0,{\rm \ \ unless \ \ }
\bigg(\sum_{k=1}^N m_k\bigg) \bmod N = 0
\label{selrule}
\ee

\ni
where $m_k,\ k=1,\ldots,N$, are highest weight components of an irrep $R$ (for
$SU(N)$, they are defined upto a common shift by an integer). Formally,
it is the confinement selection rule in low energy QCD. For example, the
fundamental representation does not obey eq. (\ref{selrule}). The first
non-trivial irrep giving non-zero answer is the adjoint. This selection
rule is called the local confinement, since two quarks cannot be
separated even by one lattice space. It is quite analogous to the one
following from the local $Z_N$ symmetry of the model \cite{Kogan&Co}.
However, the symmetry under continuous transformations (\ref{trans}) leads
to further consequences. Following to ref. \cite{BK}, let us substitute
eq. (\ref{trans}) in eq. (\ref{int}) and, as the answer cannot depend on
$\alpha$'s and $\beta$'s, we can integrate over them

\[
I_{ab}^R(\phi,\psi)=\int_{0}^{2\pi}\prod_{k=1}^N\frac{d\alpha_k
d\beta_k}{(2\pi)^2}\int  dU \:D^R_{ab}(e^{i\alpha}Ue^{i\beta})\:
e^{Ntr\phi U\psi U^+}=
\]
\be
=\sum_{c,d=1}^{d_R} P^R_{ac}I_{cd}^R(\phi,\psi)
P^R_{db}
\label{PIP}
\ee

\ni
where

\be
P^R_{ab}=\int_{0}^{2\pi}\prod_{k=1}^N\frac{d\alpha_k}{2\pi}
D^R_{ab}(e^{i\alpha})
\label{proj}
\ee

\ni
is a projector, since

\be
(P^R)^2_{ab}=\int_{0}^{2\pi}\prod_{k=1}^N\frac{d\alpha_k
d\alpha'_k}{(2\pi)^2}
D^R_{ab}(e^{i(\alpha+\alpha')})=P^R_{ab}
\label{P2eqP}
\ee

As was proven in the appendix A of ref. \cite{BK}, $P^R$ projects onto
the subspace of $V_R$ spanned by all zero-weight vectors.
This subspace, $V^{(0)}_R$, is not empty only for irreps satisfying eq.
(\ref{selrule}) and only particles corresponding to vectors from $V^{(0)}_R$
can appear even in intermediate states. From the QCD point of view,
such particles are colorless ({\em i.e.} they
coincide with their anti-particles).

As the symmetry (\ref{trans}) is
local, it cannot be spontaneously broken within the model (\ref{action}).
Details of the potential $V(\Phi)$ were not important for our
consideration.

The dimension of $V_R^{(0)}$,

\be
d^{(0)}_R=tr_{\scs{R}}P^R
=\int_{0}^{2\pi}\prod_{k=1}^N\frac{d\alpha_k}{2\pi}
\chi_{\scs{R}}(e^{i\alpha}),
\label{dim0}
\ee

\ni
increases with $N$ slower than the dimension of the whole space $V_R$,

\be
d_R=\chi_{\scs{R}}(1)
\label{dim}
\ee

\ni
When $N\to\infty$, we have

\be
d^{(0)}_R=0(\sqrt{d_R})
\ee

\ni
For example, for the adjoint

\ba
d_{adj}=N^2-1\nl
d^{(0)}_{adj}=N-1
\ea

\ni
and, hence,

\be
\langle \frac{1}{N^2} \chi_{adj}(\prod_{L} U)\rangle = 0(1/N)
\ee

\ni
for an arbitrary loop $L$ encircling a non-zero area. Such behavior is
incompatible with the usual area law.

It is clear that the
symmetry under consideration means a reduction of degrees of freedom
comparing to standard lattice QCD. On the other hand, it is this symmetry
that makes the model soluble via the Itzykson-Zuber formula
(\ref{ItZub}).
Any attempt to broke the former brokes the latter.
Therefore, at any $N$, the model is
always in the local-confinement phase, all virtual particles being
completely colorless. This picture is in agreement with direct calculations
carried out for the quadratic potential in refs. \cite{Gross,KhMak}.

\vspace{24pt}

{\bf Acknowledgements}
\vspace{12pt}

Discussions with Yu.Makeenko and M.Polikarpov are very
appreciated.

\end{document}